\documentclass[a4paper,USenglish,cleveref,autoref,thm-restate]{oasics-v2021}

\pdfoutput=1 
\hideOASIcs 
\nolinenumbers

\title{Building Trustworthy Cognitive Monitoring for Safety-Critical Human Tasks: A Phased Methodological Approach} 

\titlerunning{Building Trustworthy Cognitive Monitoring for Safety-Critical Human Tasks} 

\author{Maciej Grzeszczuk}{XR Center, Polish-Japanese Academy of Information Technology, Poland \and XR Space, Poland \and \url{https://xrc.pja.edu.pl} }{maciej.grzeszczuk@gmail.com}{https://orcid.org/0000-0002-9840-3398}{}

\author{Grzegorz Pochwatko}{Institiute of Psychology of the Polish Academy of Sciences, Poland}{}{https://orcid.org/0000-0001-8548-6916}{}

\author{Barbara Karpowicz}{XR Center, Polish-Japanese Academy of Information Technology, Poland \and XR Space, Poland}{}{https://orcid.org/0000-0002-7478-7374}{}

\author{Stanisław Knapiński}{XR Center, Polish-Japanese Academy of Information Technology, Poland \and XR Space, Poland}{}{https://orcid.org/0009-0006-0524-2545}{}

\author{Wiesław Kopeć}{XR Center, Polish-Japanese Academy of Information Technology, Poland \and XR Space, Poland}{}{https://orcid.org/0000-0001-9132-4171}{}

\authorrunning{M. Grzeszczuk et al.} 

\Copyright{Maciej Grzeszczuk and Grzegorz Pochwatko and Barbara Karpowicz and Stanisław Knapiński and Wiesław Kopeć} 


\category{} 

\relatedversion{} 

\EventEditors{John Q. Open and Joan R. Access}
\EventNoEds{2}
\EventLongTitle{42nd Conference on Very Important Topics (CVIT 2016)}
\EventShortTitle{CVIT 2016}
\EventAcronym{CVIT}
\EventYear{2016}
\EventDate{December 24--27, 2016}
\EventLocation{Little Whinging, United Kingdom}
\EventLogo{}
\SeriesVolume{42}
\ArticleNo{23}

\usepackage{graphicx}
\usepackage{todonotes}
\usepackage{subcaption}
\usepackage{hyperref}
\usepackage{xurl}
\usepackage{tabularx}
\usepackage{makecell}

\begin{document}

\maketitle

\begin{abstract}
Operators performing high-stakes, safety-critical tasks -- such as air traffic controllers, surgeons, or mission control personnel -- must maintain exceptional cognitive performance under variable and often stressful conditions. This paper presents a phased methodological approach to building cognitive monitoring systems for such environments. By integrating insights from human factors research, simulation-based training, 
 sensor technologies, and fundamental psychological principles, the proposed framework supports real-time performance assessment with minimum intrusion. The approach begins with simplified simulations and evolves towards operational contexts. Key challenges addressed include variability in workload, the effects of fatigue and stress, thus the need for adaptive monitoring for early warning support mechanisms. The methodology aims to improve situational awareness, reduce human error, and support decision-making without undermining operator autonomy. Ultimately, the work contributes to the development of resilient and transparent systems in domains where human performance is critical to safety.
\end{abstract}

\section{Introduction}

The work of an air traffic controller is often perceived by the general public as an incredibly intense job, carried out under constant tension and with full concentration. However, based on the author’s first-hand experience working GND/TWR at the country’s busiest airport, the reality is somewhat different. The job is characterized by variable intensity: periods of relatively well-organized and orderly traffic alternate with rapidly developing peaks that demand swift adaptation, heightened focus, and dynamic planning — all while remaining capable of accommodating even greater surges in case of emergencies or irregular operations.
\cite{averty2004mental,pounds2000humanfactors}. Which is not necessarily easier than constant vigilance \cite{fuenzalida2006workload}.

From a psychological standpoint, the interplay between stress and cognitive performance is crucial for understanding how air traffic controllers manage their workload. According to the classic Yerkes-Dodson law \cite{yerkes10.1002/cne.920180503,hancock10.7771/2327-2937.1024}, which posits that there is an optimal level of arousal for peak performance, both under- and over-arousal can impair decision-making and situational awareness in safety-critical environments \cite{chu10.3389/fnins.2025.1445006}. Funke et al. demonstrated that maintaining a balanced arousal level is essential for optimal performance in high-demand tasks \cite{funke10.1177/0018720816646657}. The dynamic operational environment significantly challenges not only technical skills but also psychological resilience, placing a premium on individual differences in stress regulation, attentional control, and coping strategies that profoundly influence performance outcomes \cite{bern10.1177/1071181319631106}.

Recent studies underscore the complexity of sustained attention, an essential aspect of cognitive performance. Vigilance tasks are influenced by multiple factors, including arousal levels and individual differences, suggesting that these elements should be integrated into performance assessments and interventions \cite{green10.1177/00187208221103922}. Furthermore, psychological resilience plays a critical role in navigating the stresses of high-demand tasks. Operators with better coping strategies are less likely to be adversely affected by stress, thus maintaining better performance \cite{slimani10.3390/ijerph18115885}. Integrating insights from cognitive load theory also enriches our understanding of cognitive performance. Tailored interventions designed through this lens can help sustain operator vigilance and minimize errors during peak workload periods, refining the existing cognitive monitoring systems designed for air traffic controllers \cite{visch10.1177/0276237416637959}.

Mental fatigue, defined as a state of reduced mental performance capability \cite{ICAO2016Fatigue} can build up unnoticed by the operator \cite{peukert2025subjective}, contributing as a factor reducing situational awareness (SA) \cite{endsley2017situationawareness}. Increased use of Industrial Internet of Things (IIoT) technology in ATC aims at lowering the probability of human error and improving SA \cite{viswanathan2024iiotatc}. The sensors collect Automatic Dependent Surveillance-Broadcast (ADS-B) data to clearly display aircraft position data and its callsign on the operator's display \cite{ICAO_ADS-B_Guidance_2021} in order to minimize the risk of loss of situational awareness and potential runway incursion \cite{GAPPRI2024}. Weather sensors collect and broadcast current atmospheric data \cite{EASA_SIB_2021-12R1}, essential to the flight safety, and aircraft at risk of collision automatically negotiate avoidance maneuvers \cite{EUROCONTROL_ACAS_Guide_2022}. Yet monitoring the performance of the crew or ATC personnel is not so widely adopted, with projects in the testing phase, but before production implementation \cite{raduntz2020indexing,materne2025remotetower}..

Throughout the paper, we frequently refer to examples involving air traffic controllers or aviators — partly due to the widely recognized importance of their condition and situational awareness for ensuring safety, and partly because this area offers opportunities to draw on relevant practical experience. However, the same mechanisms apply in many other areas where a human is a crucial element of a specialized process, and where safety of people or equipment depends on their quick reactions and sound decisions. This includes crane operators on construction sites, heavy machinery operators in factories, surgeons, train dispatchers, or maritime port control, but also space crew and mission control.

\section{Methods}

\subsection{Human factors}

The variability of Air Traffic Control Officer (ATCO) workload intensity mentioned in the introduction is not the only factor contributing to potential risks. The controller's work is carried out in a shift system, so it involves disruptions to the circadian cycle, limited opportunities for rest during the day (e.g. due to the fact that the rest of the family does not work according to the schedule), and finally the individual well-being of a given day, weakness, perhaps the beginnings of an illness. 

Although the ATCO is solely responsible for a traffic in his sector, the work is a teamwork - not only in terms of proper, effective communication with the aircraft, but also with other air traffic services or with vehicles moving on the airport's maneuvering area. Most of this communication is not in person - the sender is deprived from the body language, having only an intonation at his disposal. The radio he communicates with, in order to maintain compatibility with historical equipment, has a relatively low bandwidth, and commonly used amplitude modulation is susceptible to interference. For those not trained in listening to aviation talk, it is often completely incomprehensible. 

To complete the set, we must also add language problems, pronunciation, and individual accents of the speakers. The trade language of the aviation world is the aviation phraseology, based on English. But it is not colloquial English. The set of words that make up the phraseology is intended to make it difficult to misunderstand the message, e.g. the word NEGATIVE is used as a negation, but for the confirmation we use AFFIRM (not POSITIVE - so that in the case of a broken transmission it is not confused with negation, and not YES, which may be completely illegible). Other factors, such as ineffective or outdated procedures, non-ergonomic layout of the workspace, maintenance works on the infrastructure, temporary changes in the organization of airspace or weather, are additional burdens.

\subsection{Simulation environments}
\label{environments}

Experiences from a two-week experiment in a ground-based lunar facility \cite{skorupska2023casevr} indicate that even researchers directly involved in the experiment have problems with discipline in regularly completing even short reports evaluating their psychophysiological state \cite{pochwatko2023wellbeing}. Hence, the automation of such measurements, which allows determining the state of a research participant without the need to directly engage him, increases the credibility and, above all, the completeness of the collected data. But to be able to do so, we need to decide on the environment we will start our experiment in.

\begin{figure}
   \centering
   \includegraphics[width=0.9\textwidth]{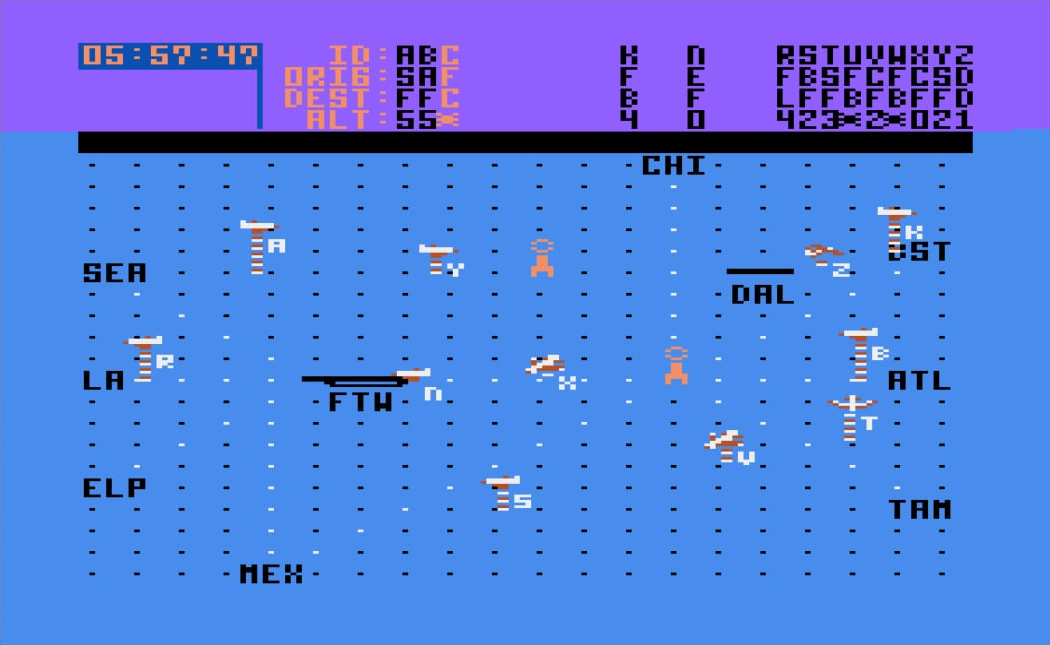}
   \caption{A screenshot from 1985 Kennedy Approach game for 8-bit Atari and Commodore 64 computers, source: \url{https://fhkd.pl/2024/12/kennedy-approach/}}
   \label{kennedy1}
\end{figure}

\begin{figure}
   \centering
   \includegraphics[width=0.9\textwidth]{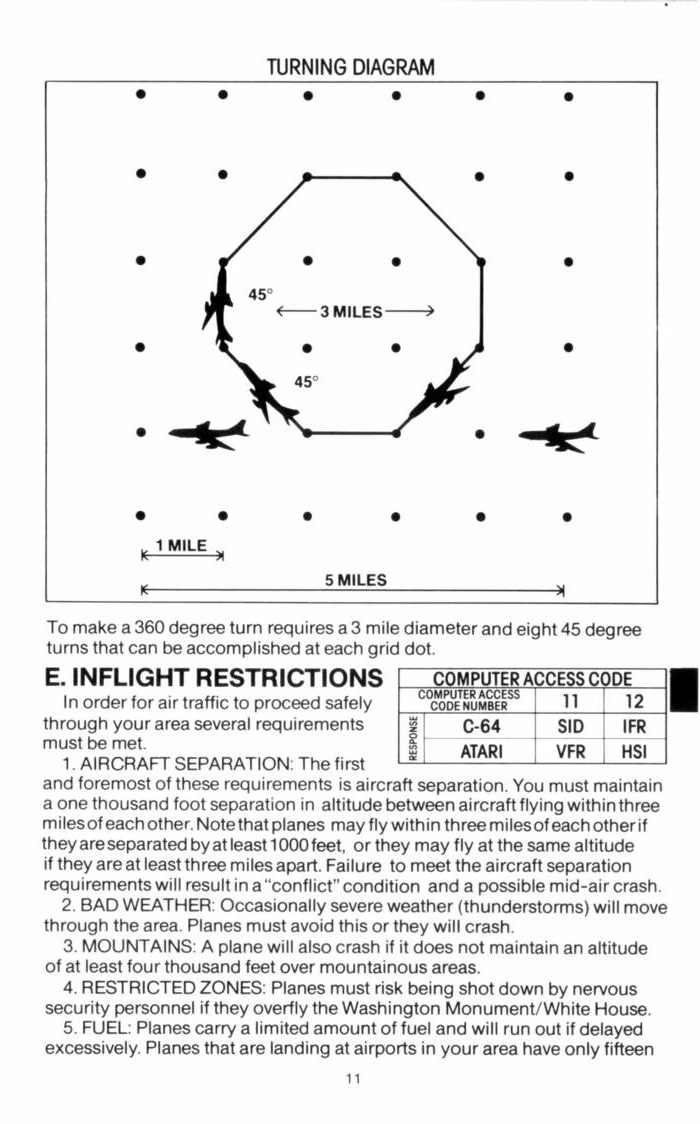}
   \caption{An excerpt from the Kennedy Approach game manual, introducing the player to the basic concepts of air traffic control. Compliance with the presented restrictions will be the basis for evaluating the performance after the session.}
   \label{kennedy2}
\end{figure}

Simulation games, which have been available on home computers since the 1980s, recreate selected aspects of reality, such as the work of a pilot or air traffic controller, in a simplified representation, lowering the entry barrier to familiarize with a given area for untrained users (see Figure \ref{kennedy1} and \ref{kennedy2}). Appropriately selected games have been found in teaching programs \cite{orhani2024gamesinkosovar, clark2015digitalgames}, not only civilian but also military, allowing to practice elements of planning and tactics of a mission without the costs and risks associated with a real endeavor \cite{reinerman2017armyaviation}. They provide the opportunity to experience the consequences of decisions made or the chosen method of implementation in the virtual world, minimizing the stress associated with the exercise or decision-making \cite{jorna1993heartrate} and offering the opportunity to test various alternative approaches in practice \cite{orhani2024gamesinkosovar}. But even though games use simplifications, the emotions they evoke in the viewer are real. The impact of their accumulation, along with fatigue and other factors that affect human performance, can be measured and recorded \cite{jorna1993heartrate}. 

\begin{figure}
   \centering
   \includegraphics[width=0.9\textwidth]{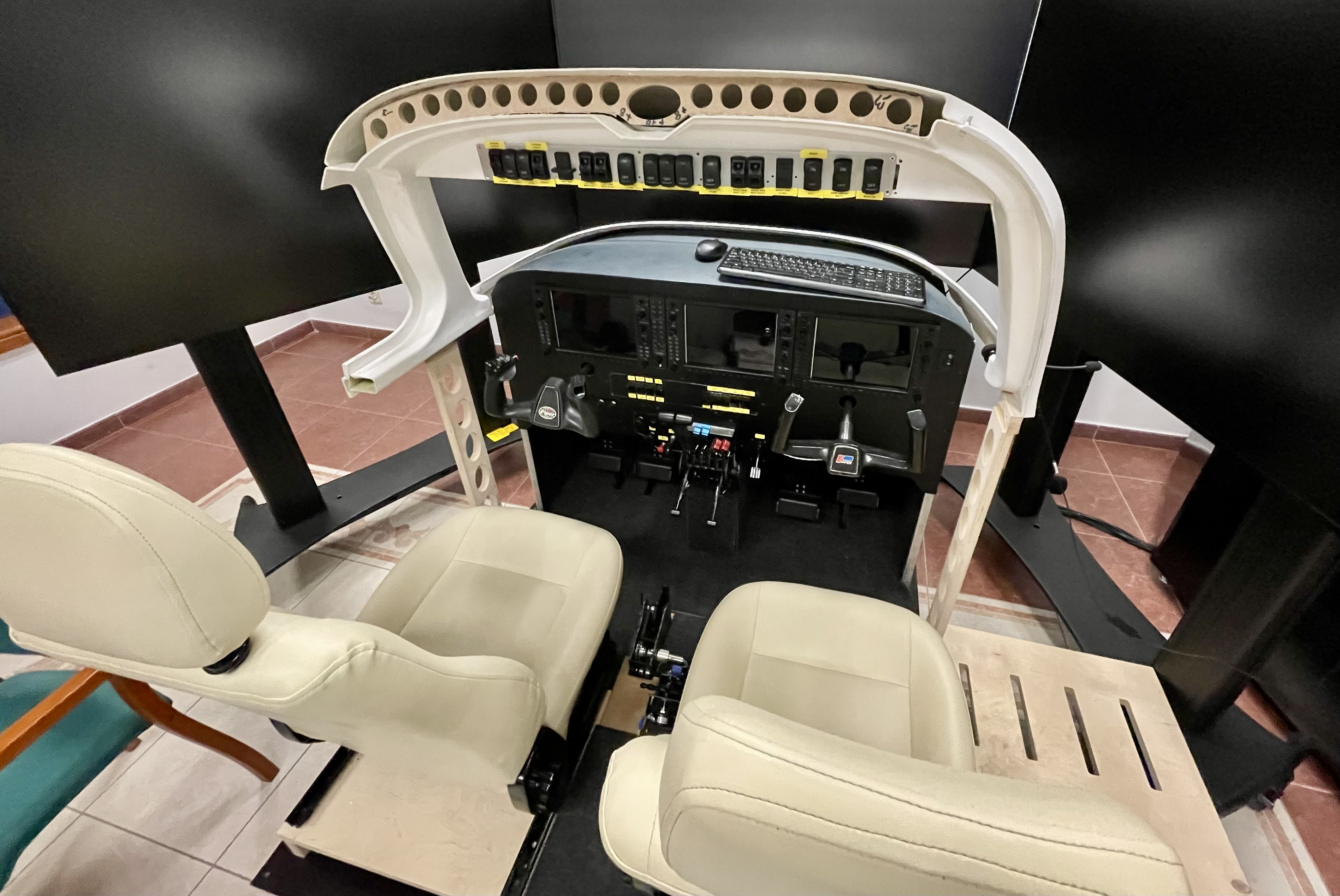}
   \caption{Flight and Navigation Procedures Trainer certified for single- and multi-engine aircraft procedures training. Source: own resources.}
   \label{seneca}
\end{figure}

\begin{figure}
   \centering
   \includegraphics[width=0.9\textwidth]{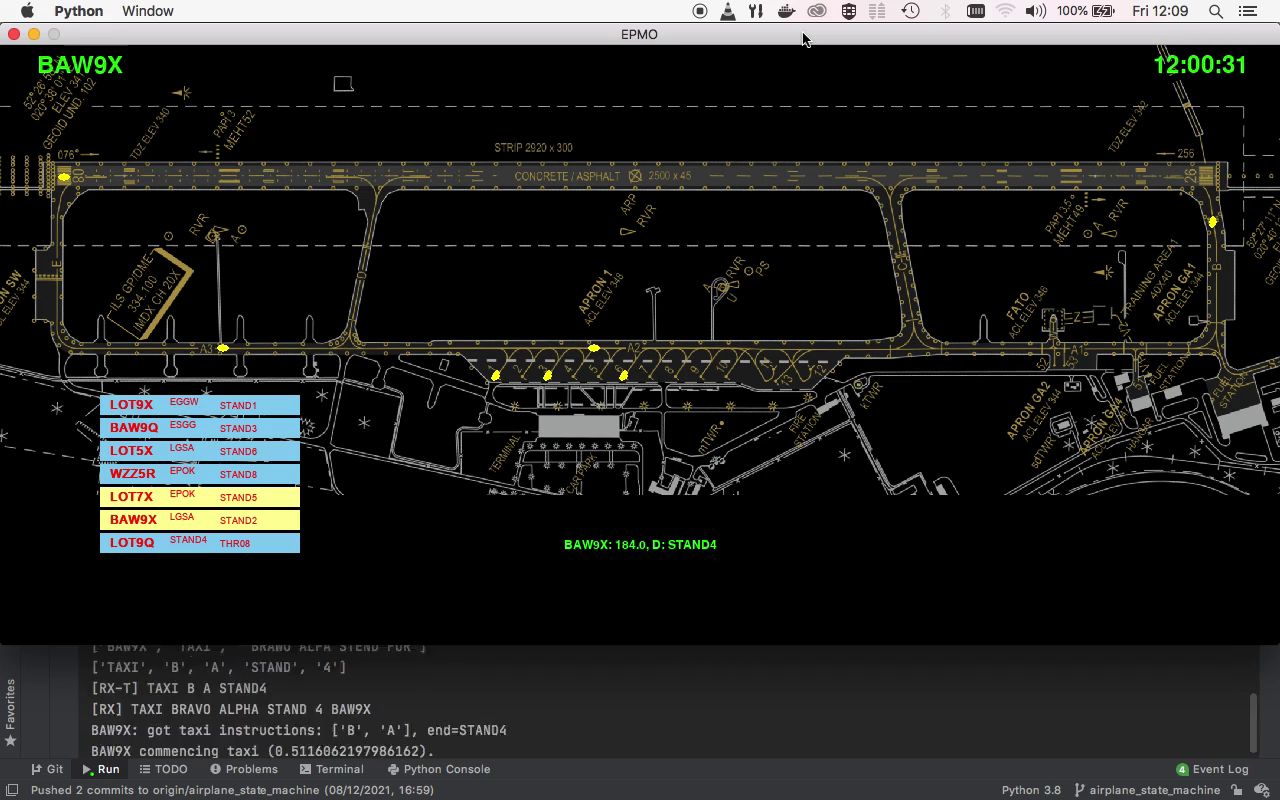}
   \caption{A screenshot from a voice-controlled custom GND simulator built in-house for the purpose of the voice sentiment analysis as a part of authors' ATCO fatigue early detection research.}
   \label{epmo}
\end{figure}

The next level is specialized simulators, which reproduce selected aspects of a given task or activity, but without full immersion. An example would be Flight and Navigation Procedures Trainers (FNPT) \cite{EASA2020fnpt} - they reproduce important parts of the cockpit of a generic aircraft class (e.g. single engine piston land aircraft - SEP(L)) and are certified by the appropriate aviation authority for use as an aid in practicing certain aspects of flight, e.g. instrument navigation and related procedures (see Figure \ref{seneca}). Alternatively, a custom solution can be created at that point (see Figure \ref{epmo}).

\begin{figure}
   \centering
   \includegraphics[width=0.9\textwidth]{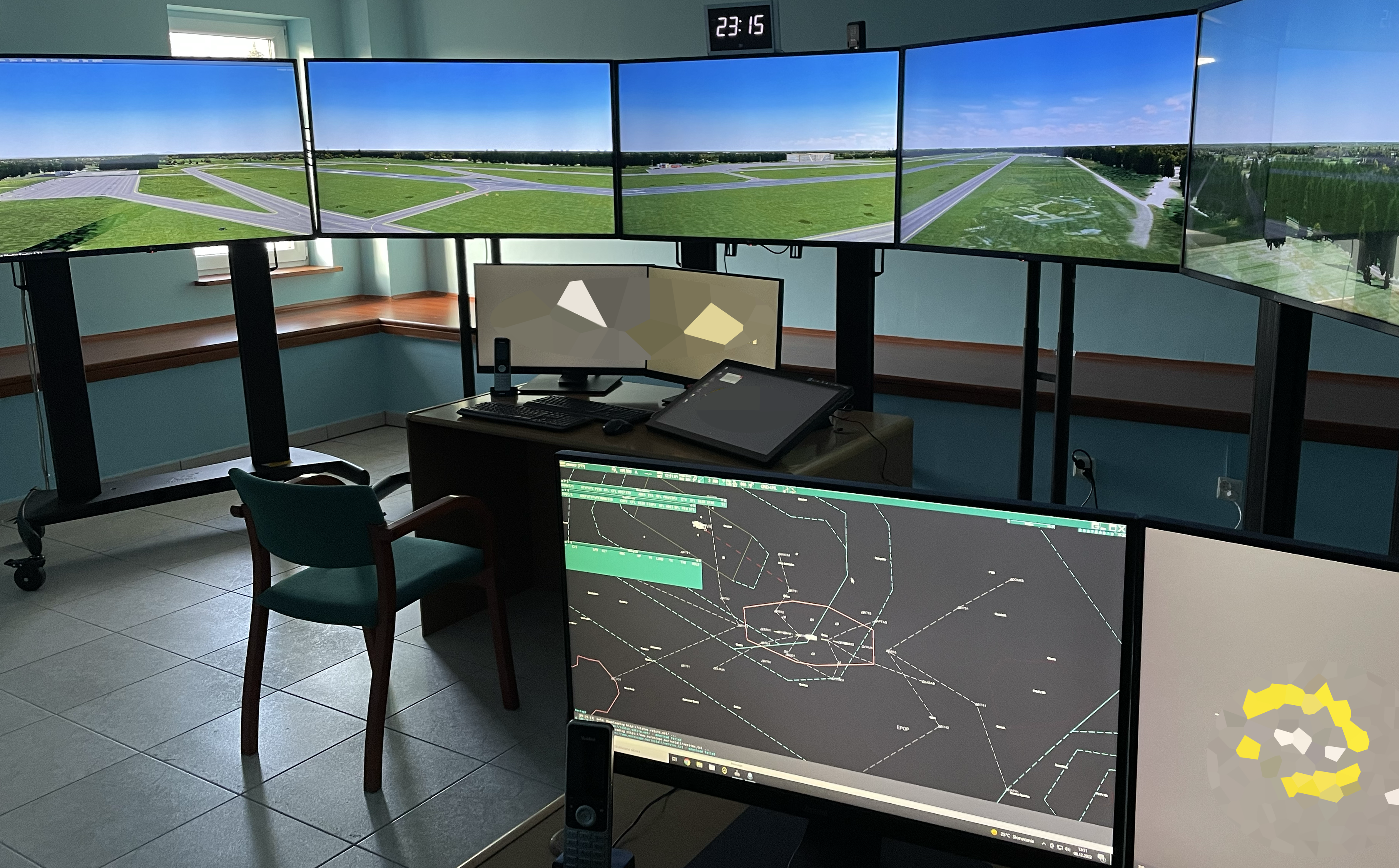}
   \caption{A certified 180-degree view tower simulator for ATCO training. In the foreground, a management station where the instructor can monitor performance in real time and interact with the simulation by initiating non-standard or emergency situations. Source: own resources.}
   \label{tower}
\end{figure}

Top tier simulation environments include the aforementioned ground-based isolation habitats, which not only reproduce the conditions inside a remote scientific base, but also the isolation and limited possibilities of leaving the habitat (by convention, or, as in the case of Antarctic bases or those built in remote areas, by practical conditions or a weather window). An example from aviation would be the Full Flight Simulator - a complete, moving cockpit of an aircraft placed on digitally-controlled actuators, with additional installations, e.g. simulating smoke and giving a deep impression of flight in its various phases. For ATCO such example would be a 360 representation of the air traffic control tower, with a live rendered view from outside the windows and fully equipped console (see Figure \ref{tower}). In order to fully reflect the dynamics of voice contact between the controller and the flight crews, real people, so-called pseudo-pilots, sit at the controls of the virtual machines and communicate with ATCO only through the simulated radio.

\subsection{Performance evaluation}

According to ICAO Annex 11, the primary objectives of Air Traffic Services (ATS), and thus of air traffic controllers, are to:

\begin{itemize}
   \item Prevent collisions (both between aircraft and between aircraft on the maneuvering area and other obstructions),
   \item Expedite and maintain an orderly flow of air traffic,
   \item Provide advice and information useful for the safe and efficient conduct of flights,
   \item Notify appropriate organization regarding aircraft in need of search and rescue aid, and assist such organizations as required.
\end{itemize}

The assessment of the first point is not binary - the mere fact that there was no collision of simulated aircraft during the exercise does not mean success. Violation of the separation criteria established for a given traffic class or space is also classified as an incident. When issuing clearances (e.g. regarding the route or cruising level), the ATCO should take into account the possible loss of communication. Therefore, aircraft should not be issue clearances that may lead to a collision. The sequence of issued clearances should cause the smallest possible delays, and if possible, traffic should flow smoothly and along routes close to optimal ones. When it comes to the readability of the message, we take into account the readability and conciseness of the message, the appropriate reaction to a possible erroneous readback as well as the appropriate rate of transmission.

The more of these conditions we are able to evaluate automatically, the easier the task of matching human parameter measurements and identifying the relationship between them and the operator's performance will be. The level of implementation of the criteria specified for the \cite{reinerman2017armyaviation} task combined with the monitoring of the subject's basic parameters (such as heart rate variability, saccadic eye movements, skin conductance, puls-monitoring infrared cameras, etc.) provide insight into the cognitive workload of the subject \cite{jorna1993heartrate,materne2025remotetower}).

\begin{table}[]
\caption{Key characteristics of simulation environments of subsequent levels of advancement.}
\label{tabelka}
\resizebox{\textwidth}{!}{%
\begin{tabular}{c|c|c|c|c}
Phase & Characteristics                                                                           & Example                                                                                 & Advantages                                                                                              & Disadvantages                                                                            \\ \hline
1     & \begin{tabular}[c]{@{}c@{}}Simplified, \\ approachable \\ by general public.\end{tabular} & A computer game.                                                                        & \begin{tabular}[c]{@{}c@{}}Inexpensive, \\ easy to recruit participants, \\ easy to learn.\end{tabular} & \begin{tabular}[c]{@{}c@{}}Too simplified, \\ wrong fit to the job.\end{tabular}         \\
2     & \begin{tabular}[c]{@{}c@{}}\\Recreates well \\ core parts of the task\end{tabular}          & \begin{tabular}[c]{@{}c@{}}Custom application, \\ FNTP II\end{tabular}                  & \begin{tabular}[c]{@{}c@{}}Good  core mechanics, \\ may be affordable\end{tabular}                      & \begin{tabular}[c]{@{}c@{}}Training needed, \\ less attractive\end{tabular}              \\
3     & Full simulation                                                                           & \begin{tabular}[c]{@{}c@{}}Full flight simulator, \\ ground based habitat.\end{tabular} & \begin{tabular}[c]{@{}c@{}}Fully immersive, \\ attractive to participants \\ (unique!)\end{tabular}     & Expensive                                                                                \\
4     & Real environment                                                                          & On-the-job tests                                                                        & Real life verification                                                                                  & \begin{tabular}[c]{@{}c@{}}Can be disruptive, \\ may require certification.\end{tabular}
\end{tabular}%
}
\end{table}

\subsection{Four-step Approach}

In our approach, we propose a sustainable development method, when we move bottom up from the simplest, approachable solution (see Table \ref{tabelka}). The first experiments on human monitoring therefore can take place on groups of volunteers who are not yet the target group of the project. Depending on who the future user will be, the choice of parameters that we want to monitor, and therefore the sensors, may differ. The environment in which it operates also matters - a different scale of integration or power efficiency of the device can be used in a stationary office environment, and another in extreme conditions (e.g. in low temperatures, carried by the operator the whole day, etc.).

When the data starts flowing, it may also turn out that some of the measurements planned to be collected is irrelevant to the task, or duplicates with the ones that are easier to collect. In such a case, it may be reasonable to trim the setup to a reasonably minimum (while ensuring user convenience and reliability).

The choice of a simple simulation environment for a start is important, even if we simplify it to the computer game environments mentioned in Section \ref{environments} - it will be easier to find volunteers who will play a potentially fun game with our sensors strapped on, than to conduct a boring study. Additionally, the initial briefing can be simpler for a game than for a complex simulator. As the developed solution takes form, we can then think about gradually upgrading the environment, increasing realism.

As our solution matures, and initial elements are validated, more complex environments can be implemented. Through early experimentation and assumption pruning, later stages can focus on those parameters that have proven significant.

Approaching the target group with a more streamlined solution later in a game has additional advantages. Technological change often causes resistance from end users, whether because they fear that technology will replace them or because bulky prototypes are inconvenient in everyday use and offer little in return.

\section{Discussion}
In this paper we explore the implications of cognitive monitoring systems into safety-critical environments. Both technical precision and human factors that subtly influence performance are emphasized. Our multilevel approach highlights the need for reliable sensor-based monitoring while ensuring that these systems adapt to the natural variability in operator responses.

An optimal level of arousal is essential for maintaining peak performance \cite{ahmad10.1007/s00779-020-01455-7}. Slight deviations can affect decision-making and situational awareness, particularly in dynamic settings like air traffic control. We show that monitoring systems should be designed to account for individual differences in response to workload. Data from sensors such as heart rate variability, eye movements, and skin conductance can be invaluable in adjusting system parameters in real time, ensuring that each operator's performance is supported without overwhelming them.

Recent advancements in wearable sensor technology have greatly expanded the possibilities for cognitive monitoring. With continuous tracking of biometric data, sensors now offer immediate feedback that allows system parameters to be adjusted on the fly, in response to stress-inducing situations \cite{zhang10.1177/1071181322661457}. The success of such systems depends both on the validity of measures and user-friendly design. 

The phased approach begins with the use of simplified simulation environments. Gamification of initial testing allows for a controlled setting to fine-tune the monitoring mechanisms and benefit from large, accessible samples of minimal requirements (i.e. non-professionals). As the system evolves and becomes increasingly integrated with training simulators or live environments, the need for adaptive and responsive monitoring becomes increasingly evident. The integrated technology should capture relevant metrics and  offer feedback that operators can intuitively understand and trust.

We are aware that introducing new monitoring technologies may raise concerns among users, fearing that the new tools will be used to evaluate and eliminate operators based on their constant performance monitoring \cite{chiligina2005cardiovascular}. But gradually refining the system and emphasizing transparency may make the process more acceptable. Especially since the dependencies discovered during the research - e.g. heatmaps collected during eye-tracking aimed towards fatigue monitoring - can help draw conclusions on improving the ergonomics of the workstation, providing a noticeable improvement in working conditions \cite{materne2025remotetower}. The monitoring tools should first and foremost serve as supportive aids to ensure enhanced performance and safety.

In summary, integrating cognitive monitoring in safety-critical applications depends on technical sophistication and deep understanding of human performance. By continuously refining these systems through advanced wearable sensor technologies and a step-by-step implementation process, we can develop tools that respond effectively to both environmental challenges and the subtle variations in how operators perform their tasks. This approach ultimately fosters safer and more efficient operational frameworks.



\bibliography{bibliography}

\begin{thebibliography}{10}

\bibitem{ahmad10.1007/s00779-020-01455-7}
M.~Ahmad, I.~Keller, D.~A. Robb, and K.~S. Lohan.
\newblock A framework to estimate cognitive load using physiological data.
\newblock {\em Personal and Ubiquitous Computing}, 27:2027--2041, 2020.
\newblock \href {https://doi.org/10.1007/s00779-020-01455-7} {\path{doi:10.1007/s00779-020-01455-7}}.

\bibitem{averty2004mental}
Philippe Averty, Christian Collet, André Dittmar, Sylvie Athenes, and Evelyne Vernet-Maury.
\newblock {Mental Workload in Air Traffic Control: An Index Constructed from Field Tests}.
\newblock {\em Aviation, space, and environmental medicine}, 75:333--41, 05 2004.

\bibitem{bern10.1177/1071181319631106}
Kyle~A. Bernhardt, Dmitri Poltavski, Thomas~V. Petros, and F.~Richard Ferraro.
\newblock {Differentiating Active and Passive Fatigue With the Use of Electroencephalography}.
\newblock {\em Proceedings of the Human Factors and Ergonomics Society Annual Meeting}, 2019.
\newblock \href {https://doi.org/10.1177/1071181319631106} {\path{doi:10.1177/1071181319631106}}.

\bibitem{chiligina2005cardiovascular}
Yulia Chiligina.
\newblock {Cardiovascular Responses to Workload and Cold-and-Hypoxic Exposure in Student Air Traffic Controllers}.
\newblock {\em Human Physiology}, pages 96--101, 01 2005.

\bibitem{chu10.3389/fnins.2025.1445006}
Zhangjie Chu, Rui Wang, and Tianyi Zhou.
\newblock {Modulation of Vigilance/Alertness Using Beta (30Hz) Transcranial Alternating Current Stimulation}.
\newblock {\em Frontiers in Neuroscience}, 2025.
\newblock \href {https://doi.org/10.3389/fnins.2025.1445006} {\path{doi:10.3389/fnins.2025.1445006}}.

\bibitem{clark2015digitalgames}
Douglas Clark, Emily Tanner-Smith, and Stephen Killingsworth.
\newblock {Digital Games, Design, and Learning: A Systematic Review and Meta-Analysis}.
\newblock {\em Review of Educational Research}, 86, 04 2015.
\newblock \href {https://doi.org/10.3102/0034654315582065} {\path{doi:10.3102/0034654315582065}}.

\bibitem{endsley2017situationawareness}
Mica Endsley.
\newblock {\em {Toward a Theory of Situation Awareness in Dynamic Systems}}, pages 9--42.
\newblock 07 2017.
\newblock \href {https://doi.org/10.4324/9781315087924-3} {\path{doi:10.4324/9781315087924-3}}.

\bibitem{EUROCONTROL_ACAS_Guide_2022}
{EUROCONTROL}.
\newblock {\em Airborne Collision Avoidance System (ACAS) Guide}, 4.1 edition, March 2022.
\newblock Dostęp online: \url{https://www.eurocontrol.int/publication/airborne-collision-avoidance-system-acas-guide}.
\newblock URL: \url{https://www.eurocontrol.int/publication/airborne-collision-avoidance-system-acas-guide}.

\bibitem{GAPPRI2024}
{EUROCONTROL}, {Flight Safety Foundation}, and {International Civil Aviation Organization}.
\newblock {Global Action Plan for the Prevention of Runway Incursions: Part I - Recommendations, Part II - Guidance and Explanatory Material}.
\newblock \url{https://www.eurocontrol.int/publication/global-action-plan-prevention-runway-incursions-gappri}, August 2024.
\newblock Validated by ACI World, CANSO, IATA.

\bibitem{EASA_SIB_2021-12R1}
{European Union Aviation Safety Agency}.
\newblock {Safety Information Bulletin: Use of Aeronautical Terminal Information Service by Air Traffic Services Units to Promulgate Information on Runway Surface Conditions – Global Reporting Format}.
\newblock Technical Report 2021-12R1, European Union Aviation Safety Agency, September 2021.
\newblock URL: \url{https://ad.easa.europa.eu/ad/2021-12R1}.

\bibitem{EASA2020fnpt}
{European Union Aviation Safety Agency (EASA)}.
\newblock {\em Easy Access Rules for Aeroplane Flight Simulation Training Devices (CS-FSTD(A))}, July 2020.
\newblock Issue 2.
\newblock URL: \url{https://www.easa.europa.eu/sites/default/files/dfu/easy_access_rules_for_aeroplane_flight_simulation_training_devices_csfstda_iss2.pdf}.

\bibitem{fuenzalida2006workload}
Eugenia Fuenzalida, Cheryl Beeler, and Laura Sohl.
\newblock {Workload history effects: A comparison of sudden increases and decreases on performance}.
\newblock {\em Current psychology (New Brunswick, N.J.)}, 25:8--14, 03 2006.
\newblock \href {https://doi.org/10.1007/s12144-006-1012-6} {\path{doi:10.1007/s12144-006-1012-6}}.

\bibitem{funke10.1177/0018720816646657}
Gregory~J. Funke, Joel~S. Warm, Carryl~L. Baldwin, Andre Garcia, Matthew~E. Funke, Michael~B. Dillard, Victor Finomore, Gerald Matthews, and Eric~T. Greenlee.
\newblock {The Independence and Interdependence of Coacting Observers in Regard to Performance Efficiency, Workload, and Stress in a Vigilance Task}.
\newblock {\em Human Factors the Journal of the Human Factors and Ergonomics Society}, 2016.
\newblock \href {https://doi.org/10.1177/0018720816646657} {\path{doi:10.1177/0018720816646657}}.

\bibitem{green10.1177/00187208221103922}
Eric~T. Greenlee, Patricia~R. DeLucia, and David~C. Newton.
\newblock {Driver Vigilance Decrement Is More Severe During Automated Driving Than Manual Driving}.
\newblock {\em Human Factors the Journal of the Human Factors and Ergonomics Society}, 2022.
\newblock \href {https://doi.org/10.1177/00187208221103922} {\path{doi:10.1177/00187208221103922}}.

\bibitem{hancock10.7771/2327-2937.1024}
P.~A. Hancock and J.~S. Warm.
\newblock A dynamic model of stress and sustained attention.
\newblock {\em Journal of Human Performance in Extreme Environments}, 7, 2003.
\newblock \href {https://doi.org/10.7771/2327-2937.1024} {\path{doi:10.7771/2327-2937.1024}}.

\bibitem{ICAO_ADS-B_Guidance_2021}
{International Civil Aviation Organization}.
\newblock {\em {ADS-B Implementation and Operations Guidance Document}}, 14.0 edition, August 2021.
\newblock URL: \url{https://www.icao.int/APAC/Documents/edocs/Revised%20ADS-B%20Implementation%20and%20Operations%20Guidance%20Document%20(AIGD)%20Edition%2014.pdf}.

\bibitem{ICAO2016Fatigue}
{International Civil Aviation Organization}, {International Business Aviation Council}, and {Flight Safety Foundation}.
\newblock {\em {Fatigue Management Guide for General Aviation Operators}}, 2016.
\newblock URL: \url{https://flightsafety.org/wp-content/uploads/2016/09/FM-for-GA-Ops-FINAL.pdf}.

\bibitem{jorna1993heartrate}
Peter Jorna.
\newblock {Heart Rate and Workload Variations in Actual and Simulated Flight}.
\newblock {\em Ergonomics}, 36:1043--54, 10 1993.
\newblock \href {https://doi.org/10.1080/00140139308967976} {\path{doi:10.1080/00140139308967976}}.

\bibitem{materne2025remotetower}
Leo Materne and Maik Friedrich.
\newblock {Supervision of Multiple Remote Tower Centers: Evaluating a New Air Traffic Control Interface Based on Mental Workload and Eye Tracking}.
\newblock {\em IEEE Transactions on Human-Machine Systems}, PP:1--10, 01 2025.
\newblock \href {https://doi.org/10.1109/THMS.2025.3527136} {\path{doi:10.1109/THMS.2025.3527136}}.

\bibitem{orhani2024gamesinkosovar}
Senad Orhani and Kyvete Shatri.
\newblock {Use of Computer Games in Kosovar Education}.
\newblock 5:28--42, 12 2024.

\bibitem{peukert2025subjective}
Maximilian Peukert, Lea CLAUS, and Lothar Meyer.
\newblock Subjective and objective fatigue dynamics in air traffic control.
\newblock {\em Industrial Health}, 02 2025.
\newblock \href {https://doi.org/10.2486/indhealth.2024-0206} {\path{doi:10.2486/indhealth.2024-0206}}.

\bibitem{pochwatko2023wellbeing}
Grzegorz Pochwatko, Wieslaw Kopec, Justyna Swidrak, Anna Jaskulska, Kinga~H. Skorupska, Barbara Karpowicz, Rafał Masłyk, Maciej Grzeszczuk, Steven Barnes, Paulina Borkiewicz, Paweł Kobyliński, Michał Pabiś-Orzeszyna, Robert Balas, Jagoda Lazarek, Florian Dufresne, Leonie Bensch, and Tommy Nilsson.
\newblock {Well-being in Isolation: Exploring Artistic Immersive Virtual Environments in a Simulated Lunar Habitat to Alleviate Asthenia Symptoms}.
\newblock In {\em {2023 IEEE International Symposium on Mixed and Augmented Reality (ISMAR)}}, pages 185--194, 2023.
\newblock \href {https://doi.org/10.1109/ISMAR59233.2023.00033} {\path{doi:10.1109/ISMAR59233.2023.00033}}.

\bibitem{pounds2000humanfactors}
Julia Pounds, Alfretia Scarborough, and Scott Shappell.
\newblock {A human factors analysis of air traffic control operational errors}, 01 2000.

\bibitem{raduntz2020indexing}
Thea Radüntz, Norbert Fürstenau, Thorsten Mühlhausen, and Beate Meffert.
\newblock {Indexing Mental Workload During Simulated Air Traffic Control Tasks by Means of Dual Frequency Head Maps}.
\newblock {\em Frontiers in Physiology}, 11:300, 04 2020.
\newblock \href {https://doi.org/10.3389/fphys.2020.00300} {\path{doi:10.3389/fphys.2020.00300}}.

\bibitem{reinerman2017armyaviation}
Lauren Reinerman-Jones, Martin Goodwin, B.F. Goldiez, Andrew Wismer, and Robert Crapanzano.
\newblock {Evaluating Game-Based Environments for Army Aviation Collective Training}.
\newblock 04 2017.

\bibitem{skorupska2023casevr}
Kinga Skorupska, Maciej Grzeszczuk, Anna Jaskulska, Monika Kornacka, Grzegorz Pochwatko, and Wies{\l}aw Kope{\'{c}}.
\newblock {A Case for VR Briefings: Comparing Communication in Daily Audio and VR Mission Control in a Simulated Lunar Mission}.
\newblock In Cezary Biele, Janusz Kacprzyk, Wies{\l}aw Kope{\'{c}}, Jan~W. Owsi{\'{n}}ski, Andrzej Romanowski, and Marcin Sikorski, editors, {\em {Digital Interaction and Machine Intelligence}}, pages 287--297, Cham, 2023. {Springer Nature Switzerland}.

\bibitem{slimani10.3390/ijerph18115885}
Maamer Slimani, Bianca Miarka, Hela Znazen, Wassim Moalla, Amri Hammami, Armin~H. Paravlić, and Nicola~Luigi Bragazzi.
\newblock {Effect of a Warm-Up Protocol With and Without Facemask-Use Against COVID-19 on Cognitive Function: A Pilot, Randomized Counterbalanced, Cross-Sectional Study}.
\newblock {\em International Journal of Environmental Research and Public Health}, 2021.
\newblock \href {https://doi.org/10.3390/ijerph18115885} {\path{doi:10.3390/ijerph18115885}}.

\bibitem{visch10.1177/0276237416637959}
Valentijn Visch.
\newblock {Warm-Up Your Audience Before Dancing With Them}.
\newblock {\em Empirical Studies of the Arts}, 2016.
\newblock \href {https://doi.org/10.1177/0276237416637959} {\path{doi:10.1177/0276237416637959}}.

\bibitem{viswanathan2024iiotatc}
Anamika Viswanathan and Kamalraj R.
\newblock {Harnessing Automation and IIoT in Air Traffic Control for Advanced Efficiency and Safety}.
\newblock {\em International Journal of Multidisciplinary Research in Science, Engineering and Technology}, 7:9979--9984, 05 2024.
\newblock \href {https://doi.org/10.15680/IJMRSET.2024.0705065} {\path{doi:10.15680/IJMRSET.2024.0705065}}.

\bibitem{yerkes10.1002/cne.920180503}
R.~M. Yerkes and J.~Dodson.
\newblock {The relation of strength of stimulus to rapidity of habit‐formation}.
\newblock {\em Journal of Comparative Neurology and Psychology}, 18:459--482, 1908.
\newblock \href {https://doi.org/10.1002/cne.920180503} {\path{doi:10.1002/cne.920180503}}.

\bibitem{zhang10.1177/1071181322661457}
Q.~Zhang, F.~Sasangohar, P.~Saravanan, N.~Ahmadi, T.~Nisar, V.~Danesh, and F.~Masud.
\newblock Real-time stress monitoring for intensive care unit (icu) nurses.
\newblock {\em Proceedings of the Human Factors and Ergonomics Society Annual Meeting}, 66:779--782, 2022.
\newblock \href {https://doi.org/10.1177/1071181322661457} {\path{doi:10.1177/1071181322661457}}.

\end{thebibliography}

\end{document}